# Theoretical complexity analysis of many-cores on a single chip

Ran Ginosar
EE, Technion, Israel
2011

## Abstract

When a single core is scaled up to *m* cores occupying the same chip area and executing the same (parallelizable) task, achievable speedup is $\sqrt{m}$, power is reduced by $\sqrt{m}$ and energy is reduced by *m*. Thus, many-core architectures can efficiently outperform architectures of a single core and a small-count multi-core.

## Introduction

How many cores should fit into a many-core single chip? The question must be asked with several scaling factors in mind:
  (a) Given a technology node, and given a specific chip area, then:
      a. How should the chip area be divided between cores and memory (ignoring other parts)
      b. Into how many cores should the core area be divided?
  (b) Given a technology node, how large a chip should be designed for a many-core?
  (c) Given a series of technology nodes, how does the many-core architecture scales?

The present analysis addresses only the first question (a).

## The Single Processor

Assume a single processor implemented on chip area A. In the next section, that area will be split into multiple processors, each one proportionally smaller. Following Fred Pollack's rule [1] regarding the dependence of processor performance on its silicon area, the maximum frequency at which a processor can execute may be claimed to grow proportionately to the square root of its area,

$$f_1 \in O\left(\sqrt{A}\right) \quad (1)$$

More generally, we assume that

$$f_1 \in O\left(A^\alpha\right) \quad (2)$$

for some $0 < \alpha < 1$, and for computational simplicity it is reasonable to employ $\alpha = 1/2$ (Hill and Marty [2] claim similarly and employ the square root for demonstrations, but stop short of adopting it for computations). To simplify the writing, we replace $\in O(\cdot)$ notation by equality, ignoring constants, where appropriate:

$$f_1 = \sqrt{A} \quad (3)$$

Assume CPI=1 and peak performance $\Pi$ [instructions per second] relative to the operating frequency, then

$$\Pi_1 = \sqrt{A} \quad (4)$$

Note that Fred Pollack's rule refers to performance and not to frequency, and CPI may be different than 1 in various architectures (although high-core-count many-core architectures are likely to be based on CPI close to 1). The entire analysis may be re-formulated based on performance, eliminating frequency from the discussion (including basing energy and power considerations not on frequency).

Clearly, in many architectures performance and frequency do not scale at the same rate. In general, the performance-frequency function in single core architectures can be presented qualitatively as in Figure 1. The central, linear segment reflects a simple 32-bit RISC core with CPI≈1 and changing frequency. On the right hand, frequency cannot be increased any further and higher performance is achieved by means of adding architectural features such as wider word, super-pipelining, super-scalar multiple issue, out-of-order and speculative execution, branch prediction and so on, all at the cost of increased area and power. On the opposite side on the left, performance can be limited for the sake of saving area or power by eliminating the pipeline, by reducing word size (16, 8 bits), by restricting the ISA and by converting the core into a 'dumb' processing element.

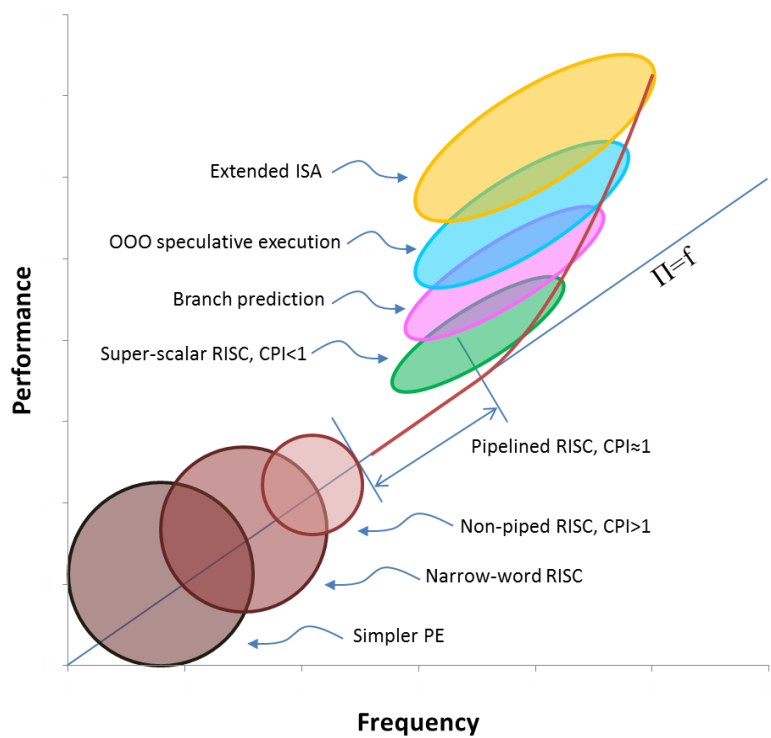

**Figure 1: Qualitative performance-frequency function on a single core**

In many-core architectures, where the area per core is rather limited, it makes sense to limit the discussion to the linear section of the chart, and thus assume that performance scales at the same rate as frequency. While it is also possible to consider the non-linear left-hand section (e.g., by studying very simple processing elements—PEs—comprising just an ALU instead of full cores) we choose to ignore it in the present analysis, which focuses on many-cores based on standard programmable core architecture rather than on many simple PEs.

Given a computational task $W$ (which is quantified on a single processor by the number of instructions that must be executed), the time to execute that task is

$$t_1 = \frac{W}{f_1} = \frac{W}{\sqrt{A}} \qquad (5)$$

Dynamic power is typically estimated as $P_D=aCV^2f$ ($a$ is activity, $C$ is capacitance of all switched nodes, $V$ is supply voltage and $f$ is operating frequency). Assume that $a$ is a given fixed factor, $C$ is proportional to area and $V$ is fixed. Then, up to a constant factor,

$$P_D = A_1 f_1 = A\sqrt{A} \tag{6}$$

Similarly, static power is proportional to area: $P_S=A$ and total power $P_1$ is (asymptotically) dominated by dynamic power. From Eqs. (5) and (6) we can compute the energy:

$$E_1 = P_1 t_1 = A\sqrt{A}\,\frac{W}{\sqrt{A}} = AW \tag{7}$$

## Ideal Plural Computing

The area A is now divided into $m$ processors, so that the area of each processor is $a=A/m$. Following Eq. (3),

$$f_1 = \sqrt{A} \qquad f_i = f_m = \sqrt{a} = \sqrt{\frac{A}{m}} \tag{8}$$

where $f_1$ is the frequency of a single processor occupying the entire area A, $f_i$ is the frequency of one of $m$ processors, and $f_m$ is the frequency of the entire $m$-processor ensemble. Similarly for performance $\Pi$,

$$\Pi_1 = \sqrt{A} \qquad \Pi_i = \sqrt{\frac{A}{m}} \qquad \Pi_m = m\sqrt{\frac{A}{m}} = \sqrt{mA} \tag{9}$$

Assume that the same computational task $W$ is divided evenly among the $m$ processors. Then each processor is assigned W/$m$ work. Assume that all $m$ processors execute simultaneously without any serial bottlenecks (for splitting the work among the $m$ processors and for merging the results) and without any overhead (when there is overhead, the total number of executed instructions exceeds W). Assume further that the frequency is reduced according to Eq. (8). This scenario is termed *Ideal Plural Computing*.

The computation time for Ideal Plural Computing (for each processor in parallel, and hence for the entire computation task) is

$$t_i = t_m = \frac{W/m}{f_i} = \frac{W}{m}\sqrt{\frac{m}{A}} = \frac{W}{\sqrt{mA}} = \frac{t_1}{\sqrt{m}} \tag{10}$$

The speedup for Ideal Plural Computing is:

$$Speedup(m) = \frac{t_1}{t_m} = \sqrt{m} \tag{11}$$

The power for Ideal Plural Computing is still proportional to the total area (A) multiplied by the frequency,

$$P_m = A\sqrt{\frac{A}{m}} = \frac{P_1}{\sqrt{m}} \tag{12}$$

Following Eqs. (10) and (12), the energy for Ideal Plural Computing is reduced linearly:

$$E_m = P_m t_m = \frac{P_1}{\sqrt{m}}\frac{t_1}{\sqrt{m}} = \frac{E_1}{m} \tag{13}$$

The Plural Computing *energydown* factor is:

$$Energydown(m) = \frac{E_1}{E_m} = m \tag{14}$$

The *ES* figure of merit combines energydown with speedup:

$$ES(m) = Energydown \times Speedup = m\sqrt{m} \tag{15}$$

Another figure of merit for Ideal Plural Computing, $ES^2$, employs the square of the speedup:

$$ES^2(m) = Energydown \times Speedup^2 = m^2 \tag{16}$$

## ET² Analysis—Single Processor

We can also analyze parallel processing from a different point of view. It was claimed in [3][4] that the cost of computation $\Theta = Et^2$ can be considered invariant for a computation task *W* using a given algorithm and architecture, over wide spectra of operating conditions that trade-off energy for computation delay. The spectra of operating conditions include voltage scaling, transistor sizing, architecture and circuit variations. The original argument in [3][4] is based on the observation that, over a wide and practical range of the supply voltage *v*, energy is proportional to $v^2$ while computation time is inversely proportional to *v*. Considering a single processor consuming $E_1$ and taking $t_1$ to compute *W*, then

$$E_1 t_1^2 = W \tag{17}$$

Considering Fig. 1 of [4] (reprinted here as Figure **2**), we assume that W reflects the lowest curve in the figure.

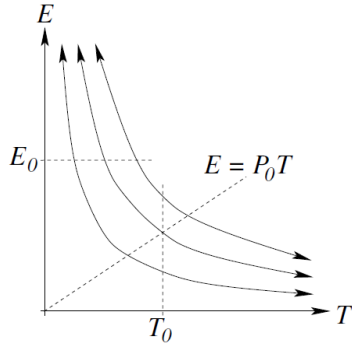

Figure 2: ET² invariance. The same problem can be solved by three different algorithms / architecures, depicted by the three different ET² charts. Only the left-most one is optimal (minimum ET²). Given energy constraint $E_0$ means selecting the leftmost point on the horizontal $E_0$ line. Given time constraint $T_0$ implies that the lowest curve yields the lowest energy solution. Given a fixed power $P_0$ constraint leads us to search the lowest point along the $E = P_0 T$ line. All three optimizations yield points along the same ET² curve. Figure 1 from [4].

Now assume that the same computation *W* is allowed longer time $t_\alpha = \alpha t_1$, $\alpha > 1$. This can be achieved by any of the trade-off parameters (e.g., voltage scaling, transistor sizing) while preserving the $Et^2$ invariance, namely selecting a point further to the right on the same $Et^2$ curve in Figure 2. Then the new energy $E_\alpha$ is $\alpha^2$ lower:

$$E_\alpha = \frac{W}{t_\alpha^2} = \frac{W}{\alpha^2 t_1^2} = \frac{E_1}{\alpha^2} \tag{18}$$

Power is reduced by an even higher factor: since $P = E/t = W/t^3$,

$$P_\alpha = \frac{W}{t_\alpha^3} = \frac{W}{\alpha^3 t_1^3} = \frac{P_1}{\alpha^3} \qquad (19)$$

Alternatively, if a smaller portion β<1 of the work is to be performed (on the same processor), we no longer expect to be on the same Et² curve so Et² now changes. The new computation can be performed, for instance, using the same parameters as above (e.g. changing neither voltage nor frequency) and so it will be completed proportionately in shorter time and less energy:

$$t_\beta = \beta t_1, \qquad E_\beta = \beta E_1 \qquad (20)$$

Then the new Et² invariant, $\Theta_\beta$, is:

$$\Theta_\beta = E_\beta t_\beta^2 = \beta^3 E_1 t_1^2 = \beta^3 W \qquad (21)$$

Note, however, that power remains unchanged, since P=E/t and both energy and time are scaled down by the same factor. Still, this is good news. Since we are allowed to trade-off energy for time, we could perform the reduced computation at the same original time $t_1$ and achieve a steeper energy saving by β³:

$$E' = \frac{\beta^3 E_1 t_1^2}{t_1^2} = \beta^3 E_1 \qquad \text{same time } t_1 \qquad (22)$$

Or we could allocate the same energy $E_1$ and further reduce time:

$$t' = \sqrt{\frac{\beta^3 E_1 t_1^2}{E_1}} = \beta^{\frac{3}{2}} t_1 \qquad \text{same energy } E_1 \qquad (23)$$

## ET² Analysis—Parallel Processor

So far we have applied the Et² model to fractional time increase and task decrease. We can now apply the results to parallelization. Reorganize the area *A* of the single processor into *m* processors and divide the work evenly among them (assuming this is possible and no overhead is required). Assuming the Ideal Plural Computing model, the frequency $f_i$ is scaled down by $\sqrt{m}$. This frequency scaling, as well as the fact that the computation per processor is now *W/m*, determine the compute time according to Eq. (10), namely $t_i = t_1/\sqrt{m}$. We can now apply Et² considerations to re-derive the same results as Eqs. (11)--(14). Note that $\beta = 1/m$ and $\Theta_\beta = \beta^3 \Theta_1$:

$$\Theta_i = \frac{1}{m^3} \Theta_1 = \frac{E_1 t_1^2}{m^3} = E_i t_i^2 = \frac{E_i t_1^2}{m}$$
$$\therefore E_i = \frac{E_1}{m^2} \qquad (24)$$

Since *m* processors now tick together,

$$E_m = m E_i = \frac{E_1}{m} \qquad (25)$$

And we reconfirm Eq. (13). Similarly,

$$P_m = m P_i = m \frac{E_i}{t_i} = m \frac{E_1}{m^2} \frac{\sqrt{m}}{t_1} = \frac{P_1}{\sqrt{m}} \qquad (26)$$

reconfirming Eq. (12). Thus, speedup and energydown are also confirmed.

We can now tabulate all parameters for the Ideal Plural Computation model. Recall that these are asymptotic measures:

|  | Single Processor | m Processors | |
|---|---|---|---|
|  |  | One core $i$ | $m$ cores ensemble |
| Area (a) | $A$ | $A/m$ | $A$ |
| Frequency (f) | $\sqrt{A}$ | $\sqrt{A/m}$ | $\sqrt{A/m}$ |
| Performance ($\Pi$) | $\sqrt{A}$ | $\sqrt{A/m}$ | $\sqrt{mA}$ |
| Compute task | $W$ | $W/m$ | $W$ |
| Compute time (t) | $W/\sqrt{A}$ | $W/\sqrt{mA}$ | $W/\sqrt{mA}$ |
| SpeedUp (SU) |  |  | $\sqrt{m}$ |
| Energy (E) | $AW$ | $AW/m^2$ | $AW/m$ |
| EnergyDown (ED) |  |  | $m$ |
| Power (P) | $A\sqrt{A}$ | $\frac{A}{m}\sqrt{\frac{A}{m}}$ | $A\sqrt{\frac{A}{m}}$ |
| PowerDown (PD) |  |  | $\sqrt{m}$ |
| Performance/Power ($\Pi P$) | $\frac{1}{A}$ | $\frac{m}{A}$ | $\frac{m}{A}$ |

# Numeric demonstration of the model

The log-log chart in Figure 3 demonstrates the relative attributes for $m=1...16K$. The total area A is set to 1M and W=1 (these are mere scaling factors that make the chart readable). The chart for the number of cores *m* is included for reference, showing a linear trend. Frequency *f* decreases with $\sqrt{m}$, as do computation time *t* and power *P*. Energy decreases linearly with *m*. Performance and speedup increase with $\sqrt{m}$, but the performance/power ratio improves linearly with *m*.

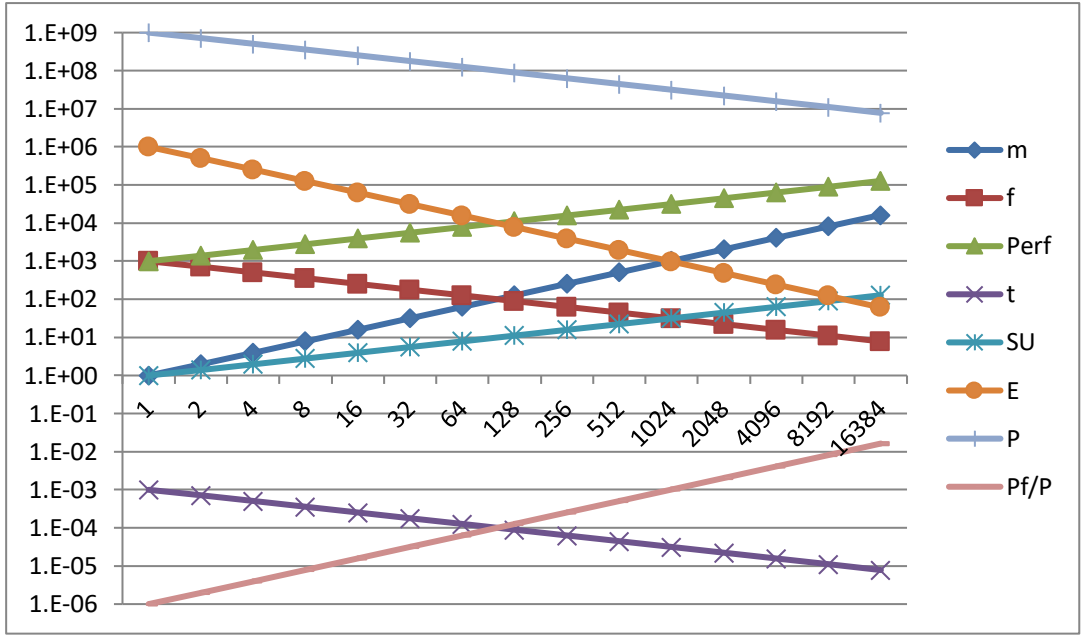

**Figure 3: Relative growth of model parameters as a function of the number of cores *m* integrated into a fixed area**

A clear characteristic of the model is that it saves power and energy in an aggressive manner. A different model could allow a fixed power budget in return for better speedup. Can such a model be constructed, and what would be the architectural justification for such a model?

# Plural Computing

Plural Computing is based on the simplified architectural model shown in Figure 4.

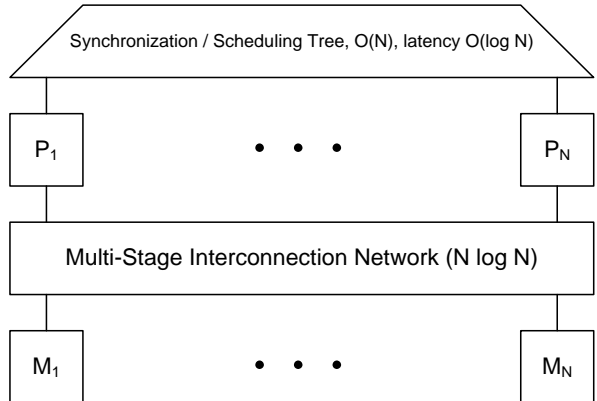

**Figure 4: Plural Computing architectural model**

The chip area A is actually divided into processors, memories, and the two networks (processors-to-memories and processors-to-scheduler). We can modify the Ideal Plural Computing model and add the cost of communications (in terms of power and area). All communications are either processor-memory (for each Read and Write) or processor-scheduler (for each end of task).

The cost of memories is less critical than the cost of communications. First, the memory area is arbitrarily subtracted from A. It may be possible to consider a desirable ratio between the number of processors $m$, the area devoted to processors and the size (area) of memories. Second, the power of memories is essentially proportional to the rate of memory access operations (read and write) and that power may be accounted for in the power for communications that facilitate memory access.

## *The Plural Programming Model*

Before considering the complexity of communications in Plural Computing, we must define how these communications are created. The programming model is based on tasks, as follows:
- All computing is organized in tasks. All code lines belong to tasks.
- All tasks execute on a shared memory. They may employ local working memory (caches and scratchpad), but its scope is task-internal and its contents disappear once a task completes.
- Precedence relations among tasks are described in a task graph, and are managed by the scheduler, which receives task completion messages and schedules subsequent tasks for execution according to the task graph.
- There are three types of tasks:
  - Singular task, which is identified by its code entry point. The scheduler initiates the execution of task $j$ on processor $i$ by sending the code entry point of $j$ to $i$.
  - Duplicable task, which is duplicated into $d$ independent concurrent instances. They can be executed in any order, including simultaneously. Each instance is identified by the code entry point (same for all *d* instances) and by its unique instance number.
- o Control task, which includes no executable code. It merely controls branch, merge and conditional points in the task graph, and is executed entirely by the scheduler.
- Tasks are not functions. They have neither arguments, nor inputs, nor outputs. Tasks only share data with other tasks via shared variables in shared memory.
- There are no synchronization points other than task completion. No BSP, no barriers.
- Memory access conflicts (simultaneous accesses to shared variables in shared memory) are resolved by the processor-to-memory network, which non-deterministically serializes conflicting accesses. There is no other access management mechanism.
- Memory access is managed according to the Concurrent Read, Exclusive Write (CREW) discipline of the PRAM model. When a task is allowed to write into a shared variable, no other task that is concurrent to it can access that variable. All concurrent tasks are allowed to read variables that are not written into by any one of them.

## Cost of Communications in Plural Computing

There are three types of communications in the Plural Computing model: Processors to the scheduler (task completion and initiation messages), processors fetching instructions, and processors accessing data (read and write).

### *Processors to scheduler*

There are one scheduler and *m* processors. Hence, the p-to-s network can be organized as a tree, and the latency is O(log *m*). However, this is irrelevant on-chip. In fact, any processor can be placed anywhere on the chip, and its distance to the scheduler is $\sqrt{A}$, the length of the chip edge. This is the conceptual measure of latency. Ideally, however, assume that latency can be hidden by creating a queue (typically of depth 1) for pre-allocation: the scheduler can send each processor the message about its next task while the present task is executing. Ideally, assume further that while *m* tasks are executing on the *m* processors, there is an ample supply of new *m* tasks ready for execution that can be pre-allocated, and that the scheduler manages to produce pre-allocations at the required rate, so that no processor waits idle for lack of work to do (and no throughput issues exist). Thus, the latency going to the scheduler is effectively zero.

The energy consumed is also proportional to the length, and must be accounted for. Assume a O(1) length scheduling message travelling the distance of $\sqrt{A}$, the dissipated energy is $\sqrt{A}$.
Next, consider power by taking the rate of scheduling messages into account. We could assume a fine-granularity model in which each task executes for time $t_t$, each processor incurs one scheduling message every $t_t$, and all *m* processors generate the rate of $m/t_t$. However, there is a simpler method for asymptotic analysis, as follows. A fine-granularity task would execute a small (roughly fixed) number of instructions. Asymptotically, the task completion rate is the same order of magnitude as the instruction execution rate. For each core that rate is the same as its frequency, and the combined rate is the same as the combined performance, $\sqrt{mA}$.

In summary,

$$E_{sched-msg} = \sqrt{A} \tag{27}$$

$$P_{sched-msg} = E_{sched-msg} \times RATE_{sched-msg} = A\sqrt{m} \tag{28}$$

### *Processors to Instructions (fetch)*

It is reasonable to assume that each processor, or a small group of processors, maintain private instruction cache. Instruction fetches are local, and their cost (in time, energy and power) may be included in the cost of executing instructions.

### *Processors to Data (read and write)*

Each access to shared memory comprises a fixed-size message that may travel across the chip, over distance $\sqrt{A}$, dissipating energy $\sqrt{A}$. Altogether, $m$ cores generate such memory access messages every $t_m$, namely at the combined rate of $m/t_m$.

In addition, the core-to-memory network consists of $m\log(m)$ switches that dissipate energy. Each of the said messages passes through $\log(m)$ switches, dissipating $\log(m)$ energy in the switches. Together (on wires and in the switches), each message dissipates $\sqrt{A} + \log(m)$ energy, at the said rate of $m/t_m$. Note that when asymptotic factors are added, scaling factors must be carefully examined.

The access rate may be estimated as follows. In typical codes, the ratio of memory accesses (load/store) to other instructions on RISC processors is fixed (e.g., about 1:5). Thus, when performing asymptotic analysis, we may assume that the memory access rate per core is the same as its frequency, and the combined rate is the same as the combined performance, $\sqrt{mA}$.

The latency to access shared memory depends on the distance, $\sqrt{A}$, and may be considered as part of the computation.

In summary,

$$E_{one-mem-access} = \sqrt{A} + \log(m) \tag{29}$$

$$P_{all-mem-access} = \left(\sqrt{A} + \log(m)\right)\sqrt{mA} \tag{30}$$

## Demonstration of the model with communications

The chart in Figure 5 shows the three power components (computing power P, scheduling power PS and memory access power PM) and their sum. While computing power decreases with *m*, communications power increases. The chart also shows the modified performance/power ratio, which flattens off at high values of *m*.

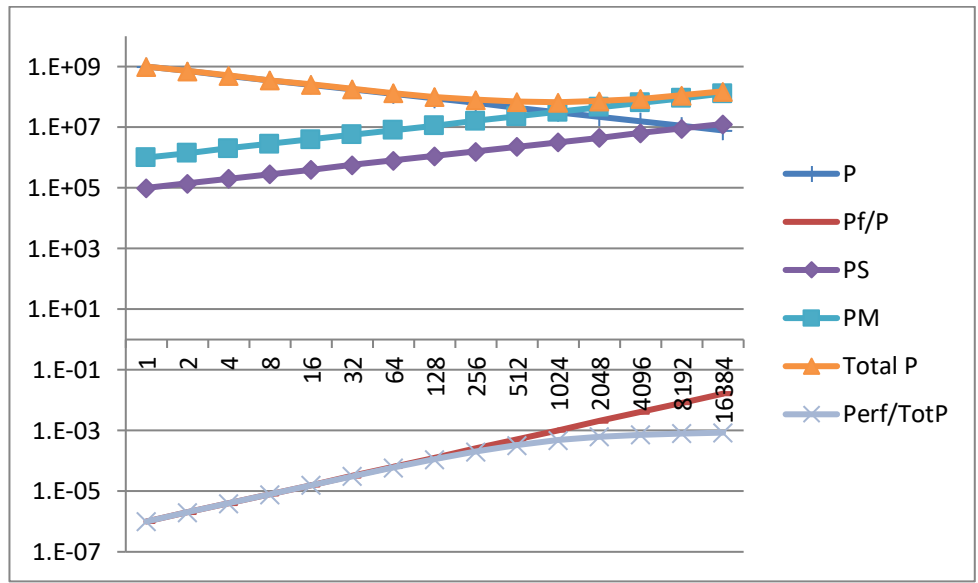

**Figure 5: Relative growth of model parameters when communications are included**